\documentclass[aps,prb,showpacs,twocolumn,floatfix]{revtex4}
\usepackage{graphicx}

\begin{document}

\title{Transient behavior of heat transport in a thermal switch}
\author{Eduardo C. Cuansing}
%\email{phyeccj@nus.edu.sg}
\author{Jian-Sheng Wang}
%\email{phywjs@nus.edu.sg}
\affiliation{Department of Physics and Centre for Computational Science 
and Engineering, National University of Singapore, Singapore 117542, 
Republic of Singapore}

\date{22 December 2009}

\begin{abstract}

We study the time-dependent transport of heat in a nanoscale thermal 
switch. The switch consists of left and right leads that are initially 
uncoupled. During switch-on the coupling between the leads is abruptly 
turned on. We use the nonequilibrium Green's function formalism and 
numerically solve the constructed Dyson equation to determine the
nonperturbative heat current. At the transient regime we find that the 
current initially flows simultaneously into both of the leads and then 
afterwards oscillates between flowing into and out of the leads. At 
later times the oscillations decay away and the current settles into 
flowing from the hotter to the colder lead. We find the transient 
behavior to be influenced by the extra energy added during switch-on. 
Such a transient behavior also exists even when there is no temperature 
difference between the leads. The current at the long-time limit
approaches the steady-state value independently calculated from the 
Landauer formula.

\end{abstract}

% 44.10.+i  <-- heat conduction
% 63.22.-m  <-- phonons or vibrational states in low-dimensional
%                  structures and nanoscale materials
% 66.70.-f  <-- nonelectronic thermal conduction and heat-pulse
%                  propagation in solids; thermal waves
% 66.70.Lm  <-- other systems such as ionic crystals,molecular
%                  crystals, nanotubes, etc.
% 05.60.Gg  <-- quantum transport

\maketitle

The physics of generation, dissipation, and manipulation of heat in 
nanoscale systems is an important topic that has recently gathered 
attention. Experiments on molecular junctions \cite{moljuncexp} found 
that the heat generated in current-carrying metal-molecule junctions 
is substantial and can affect the integrity of the device. Understanding 
how to efficiently dissipate extraneous heat in nanoscale systems is 
thus imperative in the construction of devices. In addition, the 
movement of heat may be harnessed for information processing 
\cite{infoproc}. Recent experiments have shown the viability of thermal 
transistors \cite{saira07}, thermal rectifiers using inhomogeneous 
carbon and boron nitride nanotubes \cite{chang06}, and conductance-tunable 
thermal links consisting of multiwalled carbon nanotubes \cite{chang07}. 

Most of the above-mentioned work, however, focus on the examination of
steady-state phenomena. In contrast, any physical device must function
in a time-dependent environment. Although some work has been done in
the study of time-dependent electronic transport \cite{electronic}, the
investigation of time-dependent behavior in quantum heat transport has 
not yet attracted much attention \cite{velizhanin08}. Developing 
theoretical tools and computational methods for such problems are thus 
essential to the progress of the field. In this paper we study the 
time-dependent heat current in a junction system we call a thermal 
switch. We generalize the nonequilibrium Green's function formalism 
\cite{wang08}, which is developed for steady-state situations, to the 
time-dependent case with a well-defined initial thermal state. To get 
nonperturbative results, the key steps we take are to construct and 
numerically solve a Dyson equation that does not satisfy 
time-translational invariance.

\begin{figure}[t]
\includegraphics[width=3.2in]{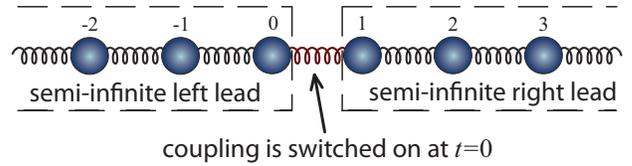}
\caption{(color online) An illustration of a quantum thermal switch.  
The labels of the first $3$ sites in each lead are shown. The coupling 
between sites $0$ and $1$ is switched on at $t = 0$.
\label{fig:switch}}
\end{figure}

Fig.~\ref{fig:switch} shows a one-dimensional chain having a coupling 
that can be switched on and off.  The semi-infinite left and right 
leads are linear chains of masses $m$. Each atom interacts with its 
left and right neighbors through an interparticle harmonic potential 
having spring constant $k$.  An on-site harmonic potential, with spring 
constant $k_0$, is also experienced by each atom. During time $t<0$ the 
left and right leads are uncoupled and are in thermal equilibrium with 
temperatures $T_{\rm L}$ and $T_{\rm R}$, respectively. At time $t = 0$ 
the coupling potential, in the form of an interparticle harmonic 
potential with the same spring constant $k$, is switched on, i.e., the 
potential between the masses labeled $0$ and $1$ in 
Fig.~\ref{fig:switch} is suddenly switched on.  We then want to know 
how the time-dependent heat current behaves in such a setup. In 
experiments on molecular junctions \cite{moljuncexp} a scanning 
tunneling microscope tip is used to stretch a molecule until a bond in 
the molecule breaks. In the thermal switch we can think of the 
switch-on as the inverse process, i.e., a bond is induced through 
proximity.

The leads follow the Hamiltonian
\begin{equation}
H^{\alpha} = \frac{1}{2} \sum_i \dot{u}_i^{\alpha} \dot{u}_i^{\alpha}
+ \frac{1}{2} \sum_{ij} u_i^{\alpha} K_{ij}^{\alpha} u_j^{\alpha},~~~
\alpha = {\rm L,R},
\end{equation}
where the sums are over all sites in the lead, the transformed 
coordinates are given by $u_i = \sqrt{m}\, x_i$, $x_i$ is the relative 
displacement of the $i$-th atom of mass $m$, and $K^{\alpha}$ is the 
spring constant matrix. The $K^{\rm L}$ and $K^{\rm R}$ matrices are 
semi-infinite tridiagonal matrices with $2k+k_0$ along the diagonal and 
$-k$ along the off-diagonal. The Hamiltonian for the switched 
coupling is 
\begin{equation}
H^{\rm LR} = \sum_{ij} u^{\rm L}_i V^{\rm LR}_{ij} u^{\rm R}_j.
\end{equation}
The coupling constant matrices $V^{\rm LR}$ and $V^{\rm RL}$ are zero 
matrices during $t<0$. After the switch-on, $V^{\rm LR}$ has one non-zero 
element $V^{\rm LR}_{01} = -k$ and $V^{\rm RL}$ has the lone non-zero 
element $V^{\rm RL}_{10} = -k$, where the matrix indices correspond to 
the labels of the masses in the leads.  Note that $H^{\rm LR} = H^{\rm RL}$. 
The time-dependent governing Hamiltonian therefore is 
$H(t) = H^{\rm L} + H^{\rm R} + H^{\rm LR}~\theta\left(t\right)$, 
where $\theta \left(t\right)$ is the Heaviside step function.

The current flowing out of the left lead
$I_{\rm L} (t) = - \left< dH^{\rm L}/dt \right>$, i.e., it is the 
expectation value of the rate of change in $H^{\rm L}$. When the switch is 
turned on the current is given by
\begin{equation}
I_{\rm L} \left( t \right) =  \hbar k~{\rm Im}\!\left[ 
\frac{\partial G^{\rm RL,<} \left(t_1,t_2\right)}{\partial t_2} 
\right]_{t_1 = t_2 = t},
\label{eq:leftcurrent}
\end{equation}
where \lq\lq${\rm Im}$\rq\rq~stands for the imaginary part. The lesser
Green's function that appears in the formula is defined as
$G^{\rm RL,<} \left(t_1,t_2\right)  =  -\frac{i}{\hbar} \bigl\langle
u^{\rm L}_0 \left(t_2\right) u^{\rm R}_1 \left(t_1\right) \bigr\rangle$,
where the subscripts $0$ and $1$ correspond to the labels of the masses.
Note that this is a two-time correlation function that does not
satisfy time-translational invariance, i.e., its time-dependence can
not be written as a difference $t_1 - t_2$. Similarly, the current 
flowing out of the right lead, 
$I_{\rm R} (t) = - \left< dH^{\rm R}/dt \right>$, when the switch is 
turned on is in the form of Eq.~(\ref{eq:leftcurrent}) except that 
the ${\rm R}$ and ${\rm L}$ superscripts are swapped. 

To determine the full, nonperturbative, current we are going to solve
the associated Dyson equation. First, we define the contour-ordered
Green's function \cite{haug96}
\begin{equation}
G^{\rm RL} \left(\tau_1,\tau_2\right) = -\frac{i}{\hbar} \left<
{\rm T}_c u^{\rm R}_1 \left(\tau_1\right) u^{\rm L}_0 
\left(\tau_2\right) \right>,
\label{eq:contour1}
\end{equation}
where the $u^{\rm R}_1$ and $u^{\rm L}_0$ are Heisenberg operators, 
${\rm T}_c$ is the contour-ordering operator, and $\tau_1$ and $\tau_2$
are complex variables on the contour $C$. To determine the current at 
time $t$ we employ a Keldysh contour $C$ that goes from time $0$ to $t$ 
and then back to $0$. Since the left and right leads are uncorrelated 
before the switch is turned on at time $0$, the contour does not have a 
complex tail after it goes back to time $0$. 

Converting to the interaction picture the contour-ordered Green's 
function shown in Eq.~(\ref{eq:contour1}) becomes
\begin{equation}
G^{\rm RL} \left(\tau_1,\tau_2\right) \!=\! -\frac{i}{\hbar}\! \left<
\!{\rm T}_c e^{-\frac{i}{\hbar}\int_C d\tau' H^{\rm LR}(\tau')} u^{\rm R}_1 
\left(\tau_1\right) u^{\rm L}_0 \left(\tau_2\right) \right>.
\label{eq:contour2}
\end{equation}
A perturbative calculation can then be done by expanding the
exponential as an infinite series. We can also use this expansion
in constructing the Dyson equation. In the series, the zeroth-order
term vanishes because it does not contain the coupling potential.
Furthermore, all the even-ordered terms also vanish because there
will be an extra $u^{\rm R}$-$u^{\rm L}$ pair without a connecting 
coupling potential.  Thus, only the odd-ordered terms survive.

\begin{figure}[t]
\includegraphics[width=3in]{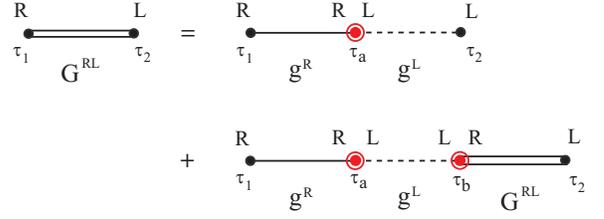}
\caption{(color online) Diagram representation of the Dyson equation. 
Each line is labelled by the Green's function it represents.  Each 
concentric dot represents a coupling vertex.
\label{fig:dyson}}
\end{figure}

We can use diagrams in constructing the Dyson equation. Shown in 
Fig.~\ref{fig:dyson} is the resulting diagram equation when we
expand the series in Eq.~(\ref{eq:contour2}). A double-line
diagram represents $G^{\rm RL}$, a single line represents the 
equilibrium Green's functions for the right lead,
$g^{\rm R} \left(\tau_1,\tau_2\right) = -\frac{i}{\hbar}
\left< {\rm T}_c u^{\rm R}_1 \left(\tau_1\right) u^{\rm R}_1
\left(\tau_2\right) \right>_0$, and a dashed line represents
the equilibrium Green's functions for the left lead,
$g^{\rm L} \left(\tau_1,\tau_2\right) = -\frac{i}{\hbar}
\left< {\rm T}_c u^{\rm L}_0 \left(\tau_1\right) u^{\rm L}_0
\left(\tau_2\right) \right>_0$. The subscript $0$ implies 
that the average is taken with respect to equilibrium distributions 
that are maintained when $t<0$ before the switch-on.

Rewriting the diagram equation in Fig.~\ref{fig:dyson} as contour
integrals, we have
\begin{eqnarray}
G^{\rm RL} \left(\tau_1,\tau_2\right) & = & \int_C 
d\tau_a~ g^{\rm R} \left(\tau_1,\tau_a\right) V^{\rm RL}~
g^{\rm L} \left(\tau_a,\tau_2\right) \nonumber \\
& & +~\int_C d\tau_a \int_C d\tau_b~ g^{\rm R} 
\left(\tau_1,\tau_a\right) V^{\rm RL} \nonumber \\
& & \times~g^{\rm L} \left(\tau_a,\tau_b\right) V^{\rm LR}~ 
G^{\rm RL} \left(\tau_b,\tau_2\right).
\label{eq:dyson1}
\end{eqnarray}
Applying Langreth's theorem to Eq.~(\ref{eq:dyson1}) and then 
iterating \cite{haug96}, we should obtain an expression for 
$G^{\rm RL,<}$. To calculate the current in Eq.~(\ref{eq:leftcurrent}) 
we need the time derivative of $G^{\rm RL,<}$, and so we differentiate 
it to get
\begin{eqnarray}
\frac{\partial G^{\rm RL,<} \left(t_1,t_2\right)}{\partial t_2}
& = & \frac{\partial G^{\rm RL,<}_1 \left(t_1,t_2\right)}
{\partial t_2} \nonumber \\
& - & k \int^t_0 dt_a~G^{\rm RL,r} \left(t_1,t_a\right)
\frac{\partial G^{\rm RL,<}_1 \left(t_a,t_2\right)}
{\partial t_2} \nonumber \\
& - & k \int^t_0 dt_a~G^{\rm RL,<}_1 \left(t_1,t_a\right)
\frac{\partial G^{\rm RL,a} \left(t_a,t_2\right)}
{\partial t_2} \nonumber \\
& + & k^2 \int^t_0 dt_a \int^t_0 dt_b~G^{\rm RL,r}
\left(t_1,t_a\right) \nonumber \\
& \times & G^{\rm RL,<}_1 \left(t_a,t_b\right)
\frac{\partial G^{\rm RL,a} \left(t_b,t_2\right)}{\partial t_2},
\label{eq:dGdt}
\end{eqnarray}
where
\begin{eqnarray}
G^{\rm RL,<}_1 \left(t_1,t_2\right) & = & -k \int^t_0 dt_a
\left\{ g^{\rm R,r} \left(t_1-t_a\right)~g^{\rm L,<}
\left(t_a-t_2\right) \right. \nonumber \\
& & +~\left. g^{\rm R,<} \left(t_1-t_a\right)~g^{\rm L,a}
\left(t_a-t_2\right) \right\}
\label{eq:1storder}
\end{eqnarray}
is the first-order term in the perturbation series in 
Eq.~(\ref{eq:contour2}). The analytic expressions for the 
equilibrium surface Green's functions $g^{\rm R,r}$, $g^{\rm R,<}$, 
$g^{\rm L,<}$, and $g^{\rm L,a}$ are known in frequency space 
\cite{wang07}. To determine their time-dependence we numerically 
calculate their corresponding Fourier transforms.

The other unknowns in Eq.~(\ref{eq:dGdt}) involve the retarded and
advanced versions of the full Green's function. We can apply 
Langreth's theorem again to Eq.~(\ref{eq:dyson1}) to determine 
expressions for these unknowns. We get
\begin{eqnarray}
G^{\rm RL,\beta} \left(t_1,t_2\right) & = & -k \int^t_0 dt_a~
G^{\rm RL,\beta}_1 \left(t_1,t_a\right)
G^{\rm RL,\beta} \left(t_a,t_2\right) \nonumber \\
& & +~G^{\rm RL,\beta}_1 \left(t_1,t_2\right),
\label{eq:retadv}
\end{eqnarray}
where $\beta = {\rm r,a}$, and the first-order term is
\begin{equation}
G^{\rm RL,\beta}_1 \left(t_1,t_2\right) = -k \int^t_0 dt_a~
g^{\rm R,\beta} \left(t_1-t_a\right)~g^{\rm L,\beta}
\left(t_a-t_2\right).
\label{eq:retadv1}
\end{equation}
To solve Eq.~(\ref{eq:retadv}) we discretize the time variable 
into $N$ segments and thus transforming the integral into a sum. 
This results in a linear problem of the form 
${\bf A}\vec{x} = \vec{b}$, where the unknown $\vec{x}$ is 
determined by performing an LU decomposition on ${\bf A}$ and then 
using $\vec{b}$ in the back-substitution. The
$\partial G^{\rm RL,a} \left(t_a,t_2\right)/\partial t_2$ term
required in Eq.~(\ref{eq:dGdt}) can also be calculated by first
differentiating Eq.~(\ref{eq:retadv}) and then finding the solution 
to the resulting equation with the time discretized. By determining 
the time derivative of the full Green's function in 
Eq.~(\ref{eq:dGdt}) the current that we calculate is a 
nonperturbative result. We follow the same steps to independently 
calculate the current flowing out of the right lead.

\begin{figure}[b]
\includegraphics[width=3.3in]{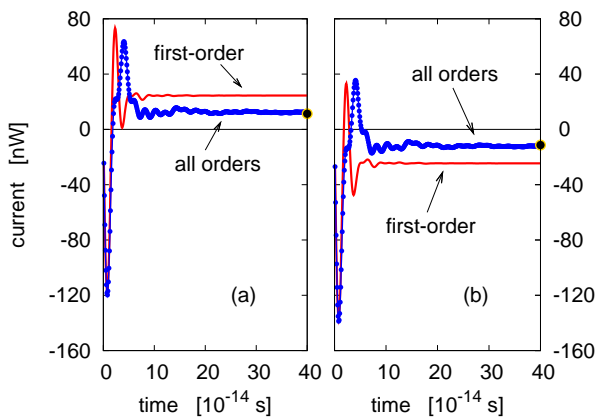}
\caption{The time-dependent current flowing out of the (a) left lead 
and (b) right lead.  The (blue online) data points are results from 
solving the Dyson equation while the (red online) line is the result 
from the first-order perturbation calculation. The average temperature 
between the leads is $T = 300~{\rm K}$. The interparticle spring 
constant is $k = 0.625$~eV/(\AA$^2$ u) while the on-site spring 
constant is $k_0 = 0.0625$~eV/(\AA$^2$ u).
\label{fig:difftemp}}
\end{figure}

Shown in Fig.~\ref{fig:difftemp} are plots of the time-dependent
current flowing out of the leads when the left and right lead
temperatures are $T_{\rm L} = 330$~K and $T_{\rm R} = 270$~K,
respectively. The average temperature is thus $T = 300$~K with
offsets of $\pm 10\%$.  The currents oscillate at a frequency
comparable to the highest phonon frequencies available in the
system and then gradually decay to their steady-state values.
The dots shown at the right edges of the plots are steady-state
values calculated independently from the Landauer formula
$I_{\rm L} = -I_{\rm R} = \frac{1}{2} \int^{\infty}_{-\infty} 
\frac{d\omega}{2\pi}~\hbar \omega \left( f_{\rm L} - f_{\rm R} \right) 
\tilde{\theta} \left(\omega\right)$, where $f_{\rm L}$ and $f_{\rm R}$ 
are the Bose-Einstein distributions of the left and right leads, 
respectively, and $\tilde{\theta} \left(\omega\right)$ is $1$ within 
the phonon band, $k_0 < \omega^2 < 4k + k_0$, and $0$ otherwise 
\cite{wang07}. At steady-state, heat should flow from the hotter to the 
colder lead, i.e., from the left to the right lead. Thus, the sign of 
the current flowing out of the left lead should be positive and for
the right lead negative. However, during the transient time, the current 
can flow in unexpected directions. Just after the switch-on, the current 
actually does not flow from the hotter to the colder lead. In 
Fig.~\ref{fig:difftemp} we see the current to flow simultaneously into 
both of the leads. There appears to be an energy source in-between the 
leads that supply the current. Recall that there is no coupling between 
the leads before the switch-on. By turning on the switch we actually add 
energy, in the form of the switched coupling potential, to the system. 
This added coupling energy supplies the current that flows into both 
leads. In Fig.~\ref{fig:difftemp} we also compare the results from 
first-order perturbation to the results from nonperturbative calculations. 
Unlike the perturbative results, nonperturbative results approach the 
steady-state values at later times. Since the switched coupling has the 
same strength as the interparticle potential we indeed expect corrections 
to perturbative calculations to be significant.

\begin{figure}[b]
\includegraphics[width=3.3in]{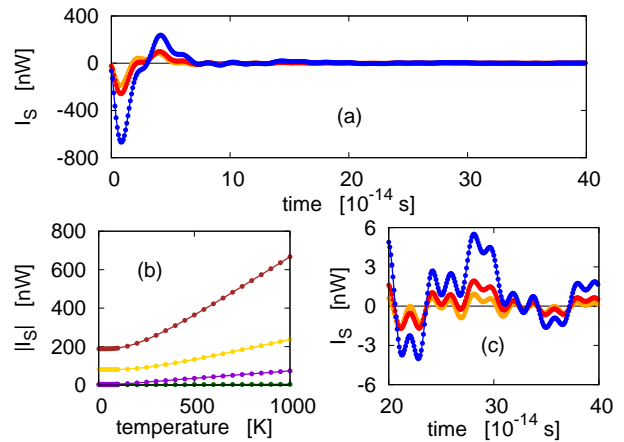}
\caption{(a) Plots of $I_{\rm S} = I_{\rm L} + I_{\rm R}$ as functions of 
time when $T = 10$~K (orange online), $T = 300$~K (red online), and 
$T = 1000$~K (blue online). The temperature offsets are $\pm 10\%$. 
(b) Plots of $\left|I_{\rm S}\right|$ as functions of $T$ at time 
$t = 0.8~[t]$ (brown online), $t = 4.1~[t]$ (yellow online), 
$t = 6.0~[t]$ (violet online), and $t = 23.0~[t]$ (green online), where
$[t] = 10^{-14}$~s. (c) An enlarged view of (a) for time $t = 20~[t]$
to $t = 40~[t]$.
\label{fig:sum}}
\end{figure}

Shown in Fig.~\ref{fig:sum} are plots of the sum of the currents,
$I_{\rm S} = I_{\rm L} + I_{\rm R}$, as functions of the time and the 
average temperature of the leads. $I_{\rm S}$ fluctuates around zero
with a fluctuation amplitude that decays with time. Fig.~\ref{fig:sum}(b) 
shows $I_{\rm S}$ to vary strongly with temperature at the transient
regime. As time goes on $I_{\rm S}$ slowly loses its dependence on 
temperature. Fig.~\ref{fig:sum}(c) shows that at later times
fluctuations in $I_{\rm S}$ still appear but are significantly smaller
than those at the transient regime. We do expect that at the steady 
state, of which the long-time limit of our data approaches, all of the 
current flowing out of the left lead should flow into the right lead and 
thus resulting in $I_{\rm S} = 0$, regardless of the value of the 
temperature. However, since energy is added during the switch-on, this 
extra energy influences the transient behavior of the system until it 
eventually dissipates out to the heat baths. Fitting the envelope to
an exponential function we find a characteristic decay time of about
$10 \times 10^{-14}$~s. At any particular time $t$ we can determine the 
energy the system absorbs or emits from 
$\left<H^{\rm LR} \left(t\right)\right> = \int_0^t 
\left(I_{\rm L} + I_{\rm R}\right) dt$.

\begin{figure}[t]
\includegraphics[width=3.3in,height=2.2in]{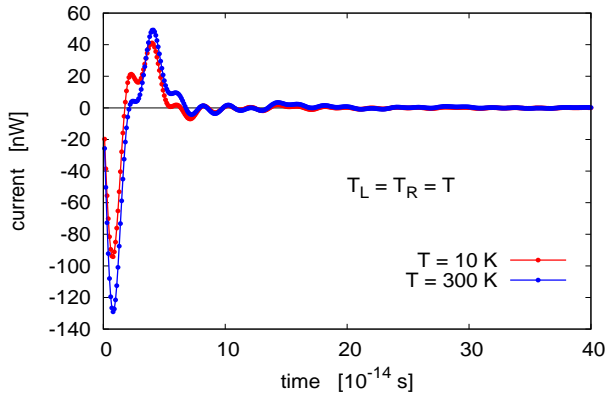}
\caption{The current flowing through the junction when the left and 
right leads have the same temperature. $T = 10$~K (red online) and 
$T = 300$~K (blue online).
\label{fig:same}}
\end{figure}

Suppose we set the temperatures of the leads to be the same. Shown in
Fig.~\ref{fig:same} is the time-dependent current, in either lead 
since the leads are indistinguishable, for such a situation. At the
steady state there should be no current flowing within the system.
However, since energy is added to the system during the switch-on, at 
the transient regime we see a fluctuating current with 
temperature-dependent amplitude flowing within the system.

The method can also be generalized in a straightforward manner to 
deal with a time-varying coupling that is on during $t > 0$. For a mildly 
increasing coupling such as, for example, $k(t) = k \tanh(ft)$, the 
transient current also initially flows simultaneously into both of the 
leads. This transient behavior persists because switching on the coupling 
introduces energy, that has to dissipate into the baths, into the system.

To summarize, we have shown an exact nonperturbative method to
calculate the time-dependent heat current in a thermal switch using 
nonequilibrium Green's functions. The Dyson equation is constructed
using Keldysh contours and the real-time Green's functions needed to 
calculate the current are determined by applying Langreth's theorem to 
the Dyson equation and numerically solving the equation with the 
discretized time variable. We set the strength of the switched coupling 
to be the same as the interparticle spring constant. Nonperturbative 
results are thus significantly different from perturbative ones. We 
find the transient current just after the switch-on to be influenced by 
the extra switched coupling energy. In particular, the initial reaction 
after switch-on is for the current to flow simultaneously into both the 
left and right leads. The current then oscillates with amplitude 
that decays with time. In the long-time limit the current approaches 
the expected steady-state values calculated independently from the 
Landauer formula. We also note that the theory presented here is not 
restricted only to one-dimensional chains but is applicable to any 
junction system where a thermal switch-on occurs.

We are grateful to Lifa Zhang, Jin-Wu Jiang, Meng Lee Leek, and
Jose Garcia for insightful discussions.  This work is supported 
in part by an NUS research grant number R-144-000-257-112.

% bibliography

\end{document}